\documentclass[aps,prb,twocolumn,groupedaddress]{revtex4-1}

\usepackage{graphicx} 
\usepackage{dcolumn} 
\usepackage{bm} 
\usepackage{amsmath}
\usepackage{amssymb}
\usepackage{latexsym}
\usepackage{epsfig}
\usepackage{amsbsy}
\usepackage{array}
\usepackage[colorlinks,linkcolor=blue,anchorcolor=red,citecolor=green]{hyperref}
\usepackage{color}
\usepackage{bookmark}
\usepackage{tabularx}

\begin{document}
\bibliographystyle{apsrev4-1}
	
\title{Hexagonal warping effect to Majorana zero modes at ends of superconducting vortex line in doped 3D strong topological insulators}
	
\author{Chuang Li$^{1,2}$} 
\author{Lun-Hui Hu$^{5,2}$} 
\email{l1hu@physics.ucsd.edu}
\author{Fu-Chun Zhang$^{2,3,4}$}
\email{fuchun@ucas.ac.cn}
	
\affiliation{$^1$Department of Physics, Zhejiang University, Hangzhou, Zhejiang, 310027, China}
\affiliation{$^2$Kavli Institute for Theoretical Sciences,  University of Chinese Academy of Sciences, Beijing 100190, China}
\affiliation{$^3$CAS Center for Excellence in Topological Quantum Computation, University of Chinese Academy of Science, Beijing, 100190, China}
\affiliation{$^4$Collaborative Innovation Center of Advanced Microstructures, Nanjing University, Nanjing 210093, China}
\affiliation{$^5$Department of Physics, University of California, San Diego, California 92093, USA}

\date{\today}
	
\begin{abstract}
	In a superconducting topological insulator, superconducting vortex line may trap 1-dimensional topological band with Majorana zero modes (MZMs) localized at its ends\cite{Vishwanath-PRL-2011}. In this work, we study the effect of hexagonal warping to the vortex phase transition. We carry out both analytical calculations based on a semiclassical formula and numerical calculations based on full quantum mechanics of Bogliubov de Gennes equation. We find that the hexagonal warping term enlarges the topological region of the vortex line as the chemical potential changes, and leads to the MZMs even in the absence of topological surface states.
\end{abstract}

\maketitle

\bookmarksetup{startatroot}

\section{Introduction}
A lot of progresses has been made in the pursuit of topological materials in the past decades, including topological insulators, topological semi-metals and topological superconductors\cite{CKChiu-RMP-2016,Bernevig-Book-2013,SQShen-Book-2012}.
The topological insulator (TI) is characterized by an insulating bulk but massless Dirac surface/edge states which are protected by time reversal symmetry. It has been realized in various systems, including the HgTe/CdTe and InAs/GaSb quantum wells\cite{SCZhang-Sci-2007,liu_prl_2008}, 3-dimensional Bi$_2$Se$_3$ family\cite{HJZhang-NatPhys-2009} and others\cite{ando_review_2013}.
The concepts of topological band theory have been naturally applied to superconductors (SC). Due to the particle-hole symmetry, topological superconductors (TSCs) can hold MZMs, whose conjugates are themselves and that are expected as candidates for the realization of quantum computer because of their non-Abelian statistics\cite{Nayak-RMP-2008}. Several theoretical and experimental efforts to realize MZMs have been carried out in different systems\cite{LFu-PRL-2008,Lutchyn-PRL-2010,Alicea-RPP-2012,Yazdani-Sci-2014}.

One effective approach to generate MZM is to hybridize TI thin film with conventional s-wave type-II SC, proposed by Fu and Kane\cite{LFu-PRL-2008}. When the uniform superconducting Cooper pairs permeate into the topological surface state of a 3D TI through proximity effect, localized MZMs may emerge at the two ends of the superconducting vortex line, which can be detected by scanning tunneling microscope and spin-selective Andreev reflection\cite{hTC-PRB-2016,JFJia-PRL-2016}.
The topological property of the vortex line has been theoretically studied. As for the doped TI, it has been shown that MZMs can still exist in the vortex even when the TI is doped to be metallic in the bulk, until a topological phase transition take places, so called vortex phase transition (VPT), when two MZMs at the two ends of the vortex line disappear\cite{Vishwanath-PRL-2011}. In the bulk HgTe, the hybridization between inverted and trivial bands can enlarge the doped topological region to stabilize MZMs\cite{CKChiu-PRL-2012} at ends of the vortex line. 
Experimentally, materials with topological surface states (TSSs) may become superconducting under certain conditions\cite{Hor-JPCS-2011}. More recently, a pristine TSC FeSe$_{0.5}$Te$_{0.5}$ is found in the iron-based superconductor, where the TSSs and superconductivity co-exist, MZMs have been observed inside the vortex core\cite{GXu-PRL-2016,DHong-Sci-2018,DHong-Sci-2018-2}.

In this work, we discuss the effect of symmetry breaking terms in the bulk Hamiltonians to the VPT and its corresponding MZMs inside the vortex line. As we know, in most materials for TIs, there is no continuous rotational symmetry, and only discrete rotational symmetry is preserved. Besides, chiral symmetry is usually broken. Taking these effects into account, the VPT may be largely affected. For example, the possible forms of the cubic momentum terms ($(k_x\pm ik_y)^3$) around $\Gamma$ point can be deduced by group theory, which exist in TIs, such as Bi$_2$Se$_3$, Bi$_2$Te$_3$, Sb$_2$Te$_3$, \emph{etc.}\cite{CXLiu-PRB-2010}. These terms are usually called hexagonal warping terms\cite{LFu-PRL-2009}. They deform TI's Fermi surface of both bulk states and surface states into hexagonal shapes. 
As we will discuss below, they can break either rotational symmetry or chiral symmetry or both.
Motivated by the breaking of symmetry in materials, in this work, we investigate the influence to the VPT by hexagonal warping terms in doped TIs. 
To obtain the critical chemical potential $\mu_c$ for the VPT, we firstly calculate the SU(2) Berry phase of the Fermi surface of TI, because the $\pi$-Berry phase is closely related to VPT; Secondly, we perform a Bogoliubov-de Gennes (BdG) calculation on an effective square lattice after taking the s-wave superconducting vortex line into account.
It turns out the topological range is enlarged by increasing the strength of the coefficients for the hexagonal warping terms. In addition, we also find the MZMs at the ends of the vortex line exist even when the topological surface state has merged into the bulk bands with respect to a large chemical potential.
	
The paper is organized as follows. We introduce the effective four-band model of TI, and discuss the effect of the hexagonal warping terms that break the continue rotational symmetry and chiral symmetry in Sec.~\ref{sec:Mod-Hmtn}. In Sec.~\ref{sec:BrPh}, we analyze the topological region of the superconducting vortex line for the doped TI, and discuss the SU(2) Berry phase of TI's Fermi surface to get some insight into the effect of the breaking of rotational symmetry. We calculate the low energy spectrum inside the vortex line by solving the BdG Hamiltonian in Sec.~\ref{sec:BdG-Calcu}. We discuss the relationship between the VPT and the disappearance of topological surface states in Sec.~\ref{sec:Topo-Surf-Stt}. In
Sec.~\ref{sec:Conclusion}, we give a brief summary.
	
\section{Model Hamiltonian, Symmetry and Fermi Surface}\label{sec:Mod-Hmtn}
We start with the effective $\mathbf{k}\cdot\mathbf{p}$ Hamiltonian \cite{HJZhang-NatPhys-2009,CXLiu-PRB-2010}  which describes the low energy physics of 3D strong TIs around the $\Gamma$ point. The minimal 4-by-4 Hamiltonian up to $O(k^2)$ reads \cite{Vishwanath-PRL-2011},  	
\begin{align}\label{eq:TI-Conti}
	H_0(\mathbf{k})= v \mathbf{k} \cdot \bm{\alpha} + (M -B k^2) \beta
\end{align}
where the basis are $\left\{ \vert t,\uparrow \rangle, \vert  t,\downarrow\rangle, \vert b,\uparrow\rangle, \vert b,\downarrow\rangle \right\}$; and the above matrices are defined as $\bm{\alpha}=\bm{\sigma} \otimes \tau_x$ and $\beta=\sigma_0 \otimes \tau_z$, where $\bm{\sigma}$ and $\bm{\tau}$ are Pauli matrices on spin and orbital subspace, respectively. Both $\sigma_0$ and $\tau_0$ are identity matrices, and $v$ is the Fermi velocity, $\mathbf{k}=(k_x,k_y,k_z)$ is the momentum vector, and $M,B$ are two parameters. 
In this work, we focus on the topological insulator phase by considering $MB>0$ that gives rise to the non-trivial $Z_2$ topological invariant \cite{LFu-PRB-2007,SQShen-Book-2012}.
The band dispersions of Eq.~\eqref{eq:TI-Conti} possess the rotational symmetry,
\begin{align}
	E_0^{\pm}(k)= \pm \sqrt{(vk)^2 +(M-Bk^2)^2}
\end{align}
where $k=\vert\mathbf{k}\vert$, and both $E_0^{+}(k)$ band and $E_0^{-}(k)$ band have two-fold degeneracy because the Hamiltonian in Eq.~\eqref{eq:TI-Conti} preserves both time-reversal symmetry $\Theta=i\sigma_y\mathcal{K}$ ($\mathcal{K}$ is the complex conjugate operator) and inversion symmetry $P=\tau_z$. The energy gap (inverted band gap) at $\Gamma$ point is $2\vert M\vert$. Furthermore, the Hamiltonian in Eq.~\eqref{eq:TI-Conti} has chiral symmetry $\mathcal{C}=\tau_y$, which leads to $\mathcal{C}H_0(\mathbf{k})\mathcal{C}^\dagger = -H_0(\mathbf{k})$.


Next, we consider the following two high-order terms (up to $O\left(k^3\right)$), both of which will break the in-plane rotation symmetry into the threefold rotation symmetry, \cite{HJZhang-NatPhys-2009,CXLiu-PRB-2010}
\begin{align}\label{eq:cubic-term}
\begin{split}
	H_{\text{hex}}(\mathbf{k}) &= H_{\text{hex},1} + H_{\text{hex},2}  \\
	H_{\text{hex},1} &= -\frac{R_1}{2} (k_+^3+k_-^3) \sigma_0 \tau_y \\
	H_{\text{hex},2} &= -\frac{R_2}{2} i(k_+^3-k_-^3) \sigma_z \tau_x
\end{split}
\end{align}
where $k_\pm= k_x \pm ik_y$ and $R_1, R_2$ are two parameters. In this work, we set $R_2=-2R_1$, smaller than the Fermi velocity $v$. As for the total Hamiltonian 
\begin{align}\label{eq:total-ham}
	H_{\text{TI}} = H_0 + H_{\text{hex}}
\end{align}
which is invariant under the
$C_{3z}=\exp\left(i\pi\sigma_z\tau_0/3\right)$ symmetry operator along the z-direction, namely $C_{3z} H(\mathbf{k}) C_{3z}^\dagger = H(C_{3z}^{-1}\mathbf{k})$. Due to the breaking of in-plane rotation symmetry, the projected (001) surface Hamiltonian only possesses threefold rotational symmetry,  $H_{\text{surf}}(k_x,k_y)=v'(\sigma_x k_y - \sigma_y k_x) + R'(k_+^3 + k_-^3)\sigma_z $, which is firstly proposed by Fu in Ref.~[\onlinecite{LFu-PRL-2009}] to study the hexagonal warping effect for TSSs of TIs. Besides, the $H_{\text{hex},1}$ term in Eq.~\eqref{eq:cubic-term} breaks the chiral symmetry $\mathcal{C}=\tau_y$. Therefore, the Hamiltonian in Eq.~\eqref{eq:total-ham} only possesses the threefold rotation symmetry, time-reversal symmetry and inversion symmetry. 

Moreover, it is valuable to note that $\mathcal{C}'=\sigma_z\tau_x$ can lead to $\mathcal{C}' H_{\text{TI}}(k_x,k_y,k_z) \mathcal{C}' = -H_{\text{TI}}(k_x,k_y,-k_z)$ when $R_1\neq0$ and $R_2=0$. Therefore, in the $k_z=0$ plane (and $k_z=\pm\pi$ for tight-binding model), $\mathcal{C}'$ severs as an in-plane chiral symmetry. In that case, only when both $R_1$ and $R_2$ are non-zero, the system breaks chiral symmetry completely.

\begin{figure}[!htbp]
	\centering
	\includegraphics[width=3.3in]{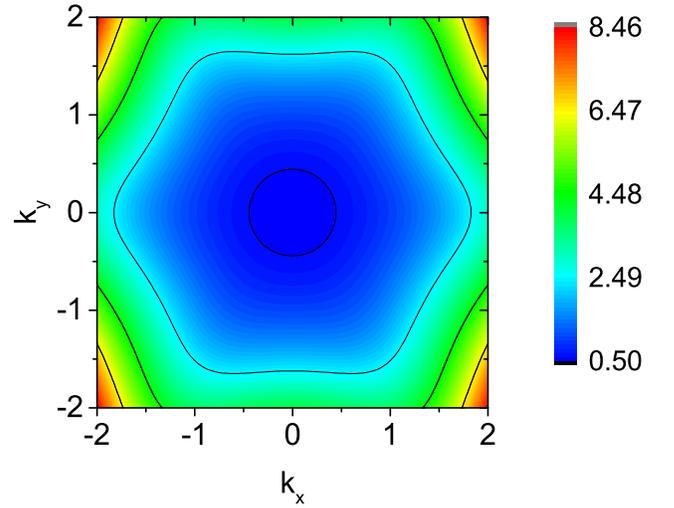}
	\caption{(color online) Represented by colors is the dispersion of TI's top band with hexagonal warping terms described by Hamiltonian Eq.~\eqref{eq:total-ham} in $k_z=0$ plane, showing the effect of Hexagonal warping terms. And it has the analytical form Eq.~\eqref{eq:total-disp-kz-0}. The parameters used here are $v=1.0$, $M=B=-0.5$, and $R_1=0.2$, $R_2=-0.4$.}
	\label{fig:Disp2-TI-Conti}
\end{figure}

To explicitly show the breaking of in-plane rotation symmetry, we investigate the band dispersions of Hamiltonian in Eq.~\eqref{eq:total-ham} in the $k_z=0$ plane,
\begin{align}\label{eq:total-disp-kz-0}
\begin{split}
	E_{\text{TI}}^{\pm}(k_\parallel) &= \pm \sqrt{(vk_\parallel)^2 +(M-B k_\parallel^2)^2 + E_{\text{hex}}} \\
	E_{\text{hex}} &= \left(R_1^2 \cos^2{3\varphi_\mathbf{k}} +R_2^2 \sin^2{3\varphi_\mathbf{k}}\right) k_\parallel^6
\end{split}
\end{align}
where $k_\parallel=\sqrt{k_x^2+k_y^2}$ and $\varphi_\mathbf{k}=\arctan{\frac{k_y}{k_x}}$. Note that each band still has two-fold degeneracy.
And the contour plot of band $E_{\text{TI}}^{+}(k_x,k_y)$ is shown in Fig.~\ref{fig:Disp2-TI-Conti}. 
In this work, we use $v=1$ and $M=B=-0.5$.
At small $k_\parallel$, $E_{\text{hex}}$ in Eq.~\eqref{eq:total-disp-kz-0} is relatively small enough to be ignored, thus the Fermi surface looks like a circle; while as $k_\parallel$ is increased, the Fermi surface has only $C_{3z}$ and inversion symmetry.

\begin{figure}[!htbp]
	\centering
	\includegraphics[width=3.3in]{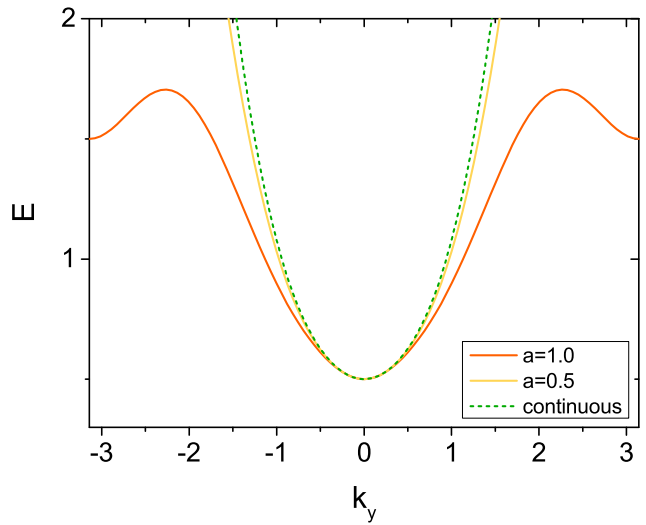}
	\caption{(color online) Comparison of dispersions at $k_x=0$ from the tight binding model with effect lattice constant $a=1$ (orange solid line) and $a=0.5$ (yellow solid line) in Eq.~\eqref{eq:total-disp-TB}, and that from the continuous model of Eq.~\eqref{eq:total-disp-kz-0} (green dashed line). The parameters used are the same as in Fig.~\ref{fig:Disp2-TI-Conti}. We will focus on the physics around $E=1.0$ where $a=1$ should capture the main physics.}
	\label{fig:Disp-TI-LatC}
\end{figure}

To simplify the numerical calculation, we can map the continue model in Eq.~\eqref{eq:total-ham} into a tight-binding (TB) model by taking the following approximations,
\begin{align}
\begin{split}
	k_i &\rightarrow \frac{1}{a} \sin(k_ia) \\
	k_i^2 &\rightarrow \frac{2}{a^2} \left\lbrack 1-\cos(k_ia)\right\rbrack \\
	k_i^3 &\rightarrow \frac{2}{a^3} \sin(k_ia) \left\lbrack 1-\cos(k_ia)\right\rbrack
\end{split}
\end{align}
where $i=x, y, z$ and $a$ is the effective lattice constant. As we will discuss below (see Sec.~\ref{sec-sub-berry}), the breaking of rotational symmetry is less important compared with the the breaking of chiral symmetry, thus the TB model is expected to capture the main physics illustrated in this work. And the TB Hamiltonian reads,
\begin{align}\label{eq:TI-TitBd}
\begin{split}
	H_\text{TI} &= H_0 + H_\text{hex}  \\
	&= \sum_i \tilde{v} \sin(k_ia) \alpha_i +\left[\tilde{M} +2\tilde{B}\sum_i \cos(k_ia) \right] \beta \\
	&\quad -\tilde{R}_1 f_1(k_x,k_y)\sigma_0 \tau_y -\tilde{R}_2 f_2(k_x,k_y)\sigma_z \tau_x
\end{split}
\end{align}
where we have defined $\tilde{v}=v/a$, $\tilde{B}=B/a^2$, $\tilde{R}_1=R_1/a^3$ and $\tilde{R}_2=R_2/a^3$.
Moreover, we have also used $\tilde{M}=M- 6\tilde{B}$, $f_1(k_x,k_y)= -4\sin(k_xa)-\sin(2k_xa) + 6 \sin(k_xa)\cos(k_ya)$, and $f_2(k_x,k_y)= -4\sin(k_ya)-\sin(2k_ya)+6\cos(k_xa)\sin(k_ya)$. 
Thus the dispersions in Eq.~\eqref{eq:total-disp-kz-0} becomes at $k_z=0$, 
\begin{align}\label{eq:total-disp-TB}
\begin{split}
  E_{\text{TI}}^{\pm}(k_x,k_y) &= \pm \Big{\lbrack}  v^2\left(\sin^2k_xa + \sin^2k_ya\right) \\
   & + \left(\tilde{M}_z +2\tilde{B}(\cos k_xa+\cos k_ya)\right)^2  \\
   & + \tilde{R}_1^2 f_1^2(k_x,k_y) +\tilde{R}_2^2 f_2^2(k_x,k_y) \Big{\rbrack}^{1/2}
\end{split}
\end{align}
Notice that $\tilde{M}_z=M-4\tilde{B}$. Then, we compare the band dispersions in the $k_z=0$ plane from the continue model (Eq.~\eqref{eq:total-disp-kz-0}) and TB model (Eq.~\eqref{eq:total-disp-TB}). The result is shown in Fig.~\ref{fig:Disp-TI-LatC} with $k_x=0$. In the $a\to0$ limit, the TB model can reproduce the continue model, indicating the validity of the above TB model in Eq.~\eqref{eq:TI-TitBd}. Because we will only focus on the interesting physics where the Fermi energy is around 1 (the inverted band gap is $|M|=0.5$), the TB model with $a=1$ should capture the main physics, and will be used here.
We have also checked that the results are similar with $a=0.5$. 	

\section{Vortex phase transition and SU(2) Berry phase}\label{sec:BrPh}
\subsection{General discussion of VPT}
In this section, we firstly review and generalize the basic concept of vortex phase transition\cite{Vishwanath-PRL-2011}. We consider the conventional s-wave even parity superconducting pairing $\Delta(\mathbf{r})i\sigma_y\tau_0$, and introduce a single vortex line into the pairing function $\Delta(\mathbf{r})$ which stretches between the top and bottom surfaces of a 3D TI. The superconducting gap may be intrinsic\cite{Hor-JPCS-2011}, and may also be induced by proximity effect\cite{LFu-PRL-2008,Yazdani-Sci-2014}. For simplicity, the 1D vortex line is oriented along the $z$ direction so that the z-direction translation symmetry is preserved. Inside the vortex line, we can study the  BdG quasi-particle spectrum of localized states. 
Then we collect all the localized 1D quasi-particle bands inside the vortex line, and judge they are topological or trivial. If the vortex line band with lowest energy is topological (trivial), we call the vortex line is topological (trivial). 
Then, we can tune some external parameters, such as the chemical potential $\mu$, to realize the topological phase transition inside vortex line, which is called vortex phase transition (VPT). 

Firstly, let us consider a trivial type-II s-wave superconductor, in which the vortex line band with lowest energy is usually fully gapped. In this case, VPT is impossible to appear. To illustrate this point, we investigate the following conventional BdG Hamiltonian\cite{Gygi-PRB-1991},
\begin{align}
  H_{\text{trivial}}(k_z) = \left( \begin{array}{cc}
                           h_0(x,y,k_z) & \Delta(x,y)e^{i\phi} \\
                           \Delta(x,y)e^{-i\phi} & -h_0^\ast(x,y,-k_z)  \\
                      \end{array} \right)
\end{align}
where $h_0(x,y,k_z)=-\frac{1}{2m_0}\left(\partial_x^2 + \partial_y^2\right) - t\cos(k_z) - E_F$ and $\Delta(x,y)\approx\Delta_0\tanh(r/\xi_0)$ with $r=\sqrt{x^2+y^2}$ and $\phi=\arctan\frac{y}{x}$. Here $m_0$ is the effective mass, $t$ is the hopping strength along the z direction, $E_F$ is the Fermi energy, $\Delta_0$ is the superconducting gap of the bulk, and $\xi_0$ is the superconducting coherent length.  In this case, the dispersion of the lowest vortex line band $E_1(k_z)$ can be approximated by\cite{deGennes-Book-1999}
\begin{align}
   E_1(k_z) \approx \Delta_0^2/(E_F+t\cos(k_z))
\end{align}
When $t\ll E_F$, it becomes $\Delta_0^2/E_F\times(1-t\cos(k_z)/E_F)$. Therefore, there exists a minimal band gap of $E_1(k_z)$ and the minimal gap is typically proportional to $\Delta_0^2/E_F$ which is usually much smaller than $\Delta_0$. Thus, there is no possibility for VPT in such trivial superconductors.

On the other hand, let us discuss the possible topological features of the vortex line. We consider the lowest vortex line band to be fully gapped, i.e., $\vert E_1(k_z)\vert >0$ for any $k_z$, so that we can define the $Z_2$ topological invariant. The vortex line is actually a 1D topological superconductor, which belongs to the class $D$ according to the Altland-Zirnbauer classification of topological phases\cite{LFu-PRB-2007,CKChiu-RMP-2016}, since only particle-hole symmetry is preserved. In analogous to the Kitaev p-wave superconducting wire\cite{Kitaev-2001-PhysU}, there exists a pair of Majorana zero modes localized at ends of the vortex line. In other words, the vortex line in this case is topologically equivalent to the Kitaev chain. 


\subsection{SU(2) Berry phase calculation}\label{sec-sub-berry}
In this subsection, we study the hexagonal warping effect to Majorana zero modes at ends of superconducting vortex line in doped 3D strong TIs. We find that the lowest vortex line band $E_1(k_z)$ is fully gapped. It becomes gapless only when the VPT takes place. Though the in-plane rotation symmetry and chiral symmetry of doped TIs is broken, Hosur's theory\cite{Vishwanath-PRL-2011} is still applicable to the model defined in Eq.~\eqref{eq:total-ham} or Eq.~\eqref{eq:TI-TitBd}.
Indeed, as firstly pointed out by Hosur in Ref.~[\onlinecite{Vishwanath-PRL-2011}], the VPT could be identified by the properties of the bulk Fermi surface. They also derived a semi-classical formula for the low energy levels to reach the critical transition point,
\begin{align}\label{eq:vpt-condition}
  E_n = \frac{\Delta_0}{l_F\xi_0} \left(2\pi n+ \pi \pm \gamma(\mu,k_z) \right)
\end{align}
where $l_F$ is the Fermi surface perimeter, $n$ is an integer and $\gamma(\mu,k_z)\in[0,2\pi]$ is the SU(2) Berry phase as a function of chemical potential $\mu$ and $k_z$.
Once either $\gamma(\mu,k_z=0)=\pi$ or $\gamma(\mu,k_z=\pi)=\pi$ are satisfied, a pair of zero modes appears and the minimal gap of vortex line bands is closed. The corresponding critical chemical potential $\mu_c$ indicates a topological phase transition since only odd number of band crossings at $k_z=0$ or $k_z=\pi$ can alter the $Z_2$ topological invariant.
In the other words, the vortex line is topological with MZMs at its ends when $\mu_c^{-}<\mu<\mu_c^{+}$; while it becomes trivial when $\mu>\mu_c^{+}$ or $\mu<\mu_c^{-}$ \cite{qin-hu-paper-2018,Vishwanath-PRL-2011,CKChiu-RMP-2016,GXu-PRL-2011}. Therefore, the $\pi$-Berry phase of bulk Fermi surfaces severs as a smoking gun to identify the topological phase transition in the 1D vortex line. Below we define a quantity,
\begin{align}\label{eq:topo-range}
  \delta \mu_c = \mu_c^{+} - \mu_c^{-} 
\end{align}
which helps to measure the topological range. Any way to enlarge $\delta \mu_c$ will play an important role in realizing and stabilizing MZMs in materials.

Next we briefly review how to calculate SU(2) Berry phase on the Fermi surface. In general, the Berry phase in momentum-space could be calculated by evaluating an integral of the Berry connection on the closed bulk Fermi surface (FS),
\begin{align}\label{eq:berry-phase}
	\gamma= \oint_{\mathbf{k}\in\text{FS}} \mathbf{A} \cdot d\mathbf{k}
\end{align}
where the Berry connection is given by,
\begin{align}\label{eq:berry-connection}
	\mathbf{A}(\mathbf{k})= i \langle \psi(\mathbf{k}) | \bm{\nabla}_\mathbf{k} | \psi(\mathbf{k}) \rangle
\end{align}
In the numerical calculation, we need to discretize both the integral in Eq.~\eqref{eq:berry-phase} and the derivation in Eq.~\eqref{eq:berry-connection}. Thus we have,
\begin{align}
\begin{split}
    e^{i\gamma} &= e^{i \sum_{\mathbf{k}} \mathbf{A} \cdot d\mathbf{k}}= e^{-\sum_{\mathbf{k}} \langle \psi(\mathbf{k}) | \bm{\nabla}_\mathbf{k} | \psi(\mathbf{k}) \rangle \cdot d\mathbf{k}} \\
     &\approx \prod_\mathbf{k} \left( 1-\langle \psi(\mathbf{k}) | \bm{\nabla}_\mathbf{k} | \psi(\mathbf{k}) \rangle \cdot d\mathbf{k} \right) \\
     &= \prod_i \left( 1-\langle \psi(\mathbf{k}_i) | \psi(\mathbf{k}_i) \rangle +\langle \psi(\mathbf{k}_i) | \psi(\mathbf{k}_{i-1}) \rangle \right) \\
     &= \prod_i \langle \psi(\mathbf{k}_i) | \psi(\mathbf{k}_{i-1}) \rangle
\end{split}
\end{align}
where all the discretized $\mathbf{k}_i$ are on the Fermi surface contour. The bulk Fermi surface studied in this work has two-fold degeneracy. So we assume the corresponding eigen-functions are $\vert\psi \rangle= \begin{pmatrix} \vert \psi_1 \rangle, \vert \psi_2 \rangle \end{pmatrix}$, where $\vert \psi_1 \rangle$ and $\vert \psi_2 \rangle$ are related to each other by the product of time-reversal and inversion symmetry. 
Then, we define the non-Abelian Berry phase $U = e^{i\gamma}$ that is just a two-by-two matrix, 
\begin{align}
    U = \prod_i \left( \begin{array}{cc}
                      \langle \psi_1(\mathbf{k}_i) | \psi_1(\mathbf{k}_{i-1}) \rangle & \langle \psi_1(\mathbf{k}_i) | \psi_2(\mathbf{k}_{i-1}) \rangle  \\ 
                      \langle \psi_2(\mathbf{k}_i) | \psi_1(\mathbf{k}_{i-1}) \rangle & \langle \psi_2(\mathbf{k}_i) | \psi_2(\mathbf{k}_{i-1}) \rangle  
                  \end{array} \right) 
\end{align}
Here $U\in$ SU(2) because of the time-reversal symmetry, so that the two eigenvalues of $U$ must be $\exp(\pm i\gamma)$ which are a pair of complex conjugates. And we fix $\gamma\in[0,2\pi]$ for simplification.

Next, we calculate and analyze the SU(2) Berry phase $\gamma(\mu,k_z)$ for the Hamiltonian in Eq.~\eqref{eq:total-ham} (continues model) and Eq.~\eqref{eq:TI-TitBd} (lattice model) which preserve both time-reversal symmetry and inversion symmetry, to study the effect of symmetry breaking (rotational symmetry and chiral symmetry), stemming from the hexagonal warping terms. 

\begin{figure}[!htbp]
	\centering
	\includegraphics[width=3.4in]{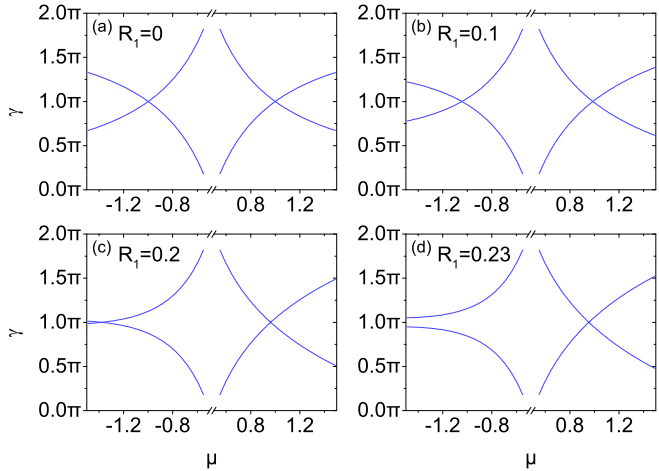}
	\caption{(color online) Berry phases around the $k_z=0$ Fermi surface (Eq.~\ref{eq:berry-phase}) using continuous model of TI. The chemical potential is set in the range $\mu\in[-1.5,-0.5]\cup[0.5,1.5]$. In (a) $R_1=0$; (b) $R_1=0.1$; (c) $R_1=0.2$; (d) $R_1=0.23$. In all cases $R_2=-2R_1$, and other parameters are the same as in Fig.~\ref{fig:Disp2-TI-Conti}. The $\mu_c^{-}$ that corresponds to the $\pi$-Berry phase is significantly increased as $R_1$ increases; while $\mu_c^{+}$ is almost unaffected. This difference is due to the breaking of chiral symmetry.}
	\label{fig:BrPh-TI-Conti}
\end{figure}

\textit{Results from the continues model.--} As we know, the symmetry breaking terms (nonzero $R_{1,2}$ in Eq.~\eqref{eq:cubic-term}) affect the bulk bands and Fermi surfaces, thus they are able to alter the Berry phases. Because the continues rotational symmetry can be only well defined for the continuous model, we firstly calculate the critical $\mu_c^\pm$ for VPT in superconducting vortex line to investigate the symmetry breaking effect in doped TIs. Firstly, we calculate the SU(2) Berry phase $\gamma(\mu,k_z=0)$ by using continue model Eq.~\eqref{eq:total-ham} with hexagonal warping term in Eq.~\eqref{eq:cubic-term}, and the results are shown in Fig.~\ref{fig:BrPh-TI-Conti}. We find that $\vert\mu_c^-\vert$ is significantly increased by increasing $R_1$, leading to the increasing of topological range $\delta\mu$. Based on Eq.~\eqref{eq:vpt-condition}, the Berry phases on Fermi surface cross $\pi$ at certain chemical potentials labeled with $\mu_c^{\pm}$, which indicates the gap closing in superconducting vortex line, thus VPT takes place. 

Next, let us discuss the relationship between $\mu_c^-$ and the symmetry breaking. There are basically three cases: (1) $R_1=R_2=0$, then both chiral symmetry and rotational symmetry are preserved. (2) Either $R_1=0$ or $R_2=0$, so rotational symmetry is broken. (3) Both $R_1$ and $R_2$ are nonzero, thus both chiral symmetry and rotational symmetry are broken.
The numerical results are summarized in Table.~\ref{tab-berry-symmetry}. Firstly, as we have addressed above, both symmetry breaking will enlarge the topological range $\delta \mu_c$. Secondly, we also notice that the breaking of chiral symmetry is dominant. This may be understood via the following argument: assume $H_0$ has chiral symmetry and $H'$ only breaks chiral symmetry, for example $H'\sim E'\tau_y$. Given that $H_0\vert\Psi_{\pm}\rangle = \pm E_0\vert \Psi_{\pm}\rangle$ and $\mathcal{C}\vert \Psi_{\pm}\rangle = \pm \vert\Psi_{\mp}\rangle$. To solve the eigen wave function of $H=H_0+H'$, namely $H\vert\Psi_{1,2}\rangle = E_{1,2}\vert \Psi_{1,2}\rangle$, to first order we find that $\vert\Psi_{1,2}\rangle \sim \vert\Psi_{\pm}\rangle + \alpha_{1,2}\vert \Psi_{\mp}\rangle$, where $\alpha_{1,2}\sim \langle \Psi_{\pm}\vert H'\vert\Psi_{\mp}\rangle\neq0$. However, if $H'$ merely breaks rotational symmetry down to $C_{3z}$ symmetry, to first order, $\alpha_{1,2}\sim \langle \Psi_{\pm}\vert H'\vert\Psi_{\mp}\rangle$ still equal to zero. Therefore, the breaking of chiral symmetry may significantly affect the eigen wave function and consequently alters the SU(2) berry phase. From this point of view, we may also argue that the lattice model should be valid when the chemical potential is small ($k_F$ is correspondingly small), through the discrete rotational symmetry of lattice model is different from the continuous model.	

\begin{table}[!htbp]
	\caption{\label{tab-berry-symmetry}Calculated critical $\mu_c^{\pm}$ that corresponds to SU(2) $\pi$-Berry phase of Fermi surface in the $k_z=0$ plane. Different $R_1$ and $R_2$ are used.}
	\begin{tabular*}{0.46\textwidth}{@{\extracolsep{\fill}}|r r|r|r|r|}
		\hline 
		$R_1$ & $R_2$ & Symmetry & $\mu_c^{-}$ & $\mu_c^{+}$ \\ 
		\hline
		0.0 & 0.0 & Chiral, Rot & -1.0 & -1.0 \\ 
		\hline 
		0.0 & -0.2 & Chiral & -1.01 & 1.01 \\ 
		\hline 
		0.0 & -0.4 & Chiral & -1.035 & 1.035 \\  
		\hline 
		0.2 & 0.0 & Chiral & -1.01 & 1.01 \\ 
		\hline 
		0.2 & -0.2 & None & -1.1 & 0.975 \\ 
		\hline 
		0.2 & -0.4 & None & -1.38 & 0.965 \\ 
		\hline 
		0.4 & 0.0 & Chiral & -1.035 & 1.035 \\ 
		\hline 
		0.4 & -0.2 & None & -1.38 & 0.965 \\ 
		\hline 
		0.4 & -0.4 & None & $<$-1.5 & 0.935 \\ 
		\hline 
	\end{tabular*} 
\end{table}

From Tab.~\ref{tab-berry-symmetry}, we also notice that the breaking of chiral symmetry only leads to the increasing of $\vert\mu_c^-\vert$ (hole doping region), while $\vert\mu_c^+\vert$ is almost unaffected. In fact, this depends on the parameters we used. For the Hamiltonian in Eq.~\eqref{eq:total-ham} or Eq.~\eqref{eq:TI-TitBd}, we find an interesting symmetry for the $k_z=0$ plane: $\mathcal{C}'^{-1} H_{\text{TI}}(-R_2) \mathcal{C}'= -H_{\text{TI}}(R_2)$ where the chiral symmetry is $\mathcal{C}'=\sigma_z\tau_x$. It means if we replace $R_2$ by $-R_2$ in the calculation, the $\pi$-Berry phase takes place at $-\mu_c^+<0$ and $-\mu_c^->0$. Therefore, the critical chemical potential in the electron doping region is enlarged.  


\begin{figure}[!htbp]
	\centering
	\includegraphics[width=3.4in]{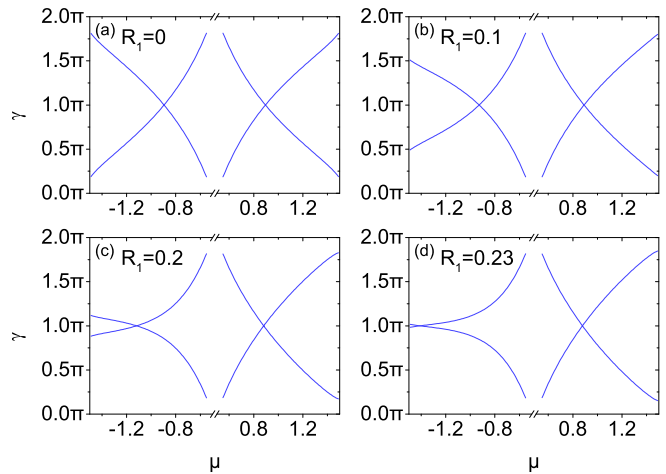}
	\caption{(color online) Berry phases around the $k_z=0$ Fermi surface (Eq.~\ref{eq:berry-phase}) using TB model of TI with effective lattice constant $a=1$. All the other parameters are the same as in the continuous model in Fig.~\ref{fig:BrPh-TI-Conti}. Similarly, as $R_1$ increases, the $\mu_c^{-}$ with $\pi$-Berry phase is significantly decreased while $\mu_c^{+}$ unaffected.}
	\label{fig:BrPh-TI-TitBd}
\end{figure}	

\textit{Results from the lattice model.--} Next, we utilize the TB Hamiltonian in Eq.~\eqref{eq:TI-TitBd} to perform the calculation of $\gamma(\mu,k_z)$ for different values of $R_{1,2}$. There exists a closed bulk Fermi surface only when the chemical potential $\vert\mu\vert>\vert M\vert=0.5$, and it has two-fold degeneracy. And our calculation shows it is only possible to hunt for the $\pi$-Berry phase of bulk Fermi surface in the $k_z=0$ plane, namely $\gamma(\mu_c^{\pm},k_z=0)=\pi$. Again, from Eq.~\eqref{eq:vpt-condition}, we already know the $\pi$-Berry phase corresponds to the gap closing of the localized band inside vortex line.
The results for Berry phase are shown in Fig.~\ref{fig:BrPh-TI-TitBd}, where we plot $\gamma$ and $-\gamma+2\pi$. The results are almost the same as that calculated from continuous model(see Fig.~\ref{fig:BrPh-TI-Conti}). And it also indicates that the $H_{\text{hex}}$ term in Eq.~\eqref{eq:cubic-term} can help to enlarge $\delta \mu_c$ defined in Eq.~\eqref{eq:topo-range} (topological range) in which topological signature persists. It can convince us that the results from the lattice model are generally reliable. 
In Fig.~\ref{fig:BrPh-TI-TitBd}(a) without hexagonal warping, we notice that $\mu_c^+=-\mu_c^-$ because of the existence of the chiral symmetry that makes  $\mathcal{C}H_0(\mathbf{k})\mathcal{C}^\dagger = -H_0(\mathbf{k})$. And $\vert \mu_c^{\pm}\vert \approx 0.90 \sim v \sqrt{M/B}$ is predicted in Ref.~[\onlinecite{Vishwanath-PRL-2011}]. In Fig.~\ref{fig:BrPh-TI-TitBd}(b-d), we increase $R_1$ from $0.1$ to $0.23$, we find $\mu_c^{+}$ is almost unchanged, say $\mu_c^{+}\approx 0.89$; while $|\mu_c^{-}|$ is enlarged. Explicitly, when $R_1=0.23$, $\mu_c^{-}=-1.41$, so that the topological range is $\delta\mu_c=2.30$, larger than that in the case with $R_1=0$ ($\delta\mu_c\approx 1.8$).

\section{BdG calculation for VPT}\label{sec:BdG-Calcu}
In this section, we provide an alternative way to identify the VPT and to study the hexagonal warping effect to VPT. As we know, the most rigorous and straightforward way to determine the VPT is just to solve the BdG Hamiltonian to hunt for the gap closing of the lowest vortex line bands, i.e., band touchings. As discussed in the above section, the lattice mode is valid in discussing the VPT. Because the chiral symmetry breaking plays an dominant role to the VPT physics, rather than the rotational symmetry.
And the vortex profile in the $x-y$ plane breaks the translation symmetry, thus we need a real space TB model. But $k_z$ is still a good quantum number. The Hamiltonian in Eq.~\eqref{eq:TI-TitBd} becomes $H_{\text{TI}} = H_0 + H_{\text{hex}}$ where $H_0$ and $H_{\text{hex}}$ are given by,
\begin{widetext}
	\begin{align}
	\begin{split}
	H_0(k_z) 
	=& \sum_{x,y} \frac{\tilde{v}}{2i} \left(C_{x+1,y,k_z}^\dagger \alpha_x C_{x,y,k_z} +C_{x,y+1,k_z}^\dagger \alpha_y C_{x,y,k_z} -\text{H.c.}\right) \\
	& +\sum_{x,y} C_{x,y,k_z}^\dagger \left\{ \tilde{v} \sin(k_za) \alpha_z +\left[ \tilde{M}+2\tilde{B}\cos(k_za) \right]\beta \right\} C_{x,y,k_z} \\
	& +\sum_{x,y} \tilde{B} \left(C_{x+1,y,k_z}^\dagger \beta C_{x,y,k_z} +C_{x,y+1,k_z}^\dagger \beta C_{x,y,k_z} +\text{H.c.}\right)
	\end{split} \\
	\begin{split}
	H_\text{hex}(k_z)
	=& \sum_{x,y} -2i\tilde{R}_1\left(C_{x+1,y,k_z}^\dagger \sigma_0 \tau_y C_{x,y,k_z}-\text{H.c.}\right) +\sum_{x,y} -i\tilde{R}_1\left(C_{x+2,y,k_z}^\dagger \sigma_0 \tau_y C_{x,y,k_z}-\text{H.c.}\right) \\
	& +\sum_{x,y} 2i\tilde{R}_1\left(C_{x+1,y+1,k_z}^\dagger \sigma_0 \tau_y C_{x,y,k_z} +C_{x+1,y-1,k_z}^\dagger \sigma_0 \tau_y C_{x,y,k_z} -\text{H.c.}\right) \\
	& +\sum_{x,y} -2i\tilde{R}_2\left(C_{x,y+1,k_z}^\dagger \sigma_z \tau_x C_{x,y,k_z}-\text{H.c.}\right) +\sum_{x,y} -i\tilde{R}_2\left(C_{x,y+2,k_z}^\dagger \sigma_z \tau_x C_{x,y,k_z}-\text{H.c.}\right) \\
	& +\sum_{x,y} 3i\tilde{R}_2\left(C_{x+1,y+1,k_z}^\dagger \sigma_z \tau_x C_{x,y,k_z} +C_{x-1,y+1,k_z}^\dagger \sigma_z \tau_x C_{x,y,k_z} -\text{H.c.}\right)
	\end{split}
	\end{align}
\end{widetext}	
where the basis for the single-particle Hamiltonian is, $\Psi_{x,y,k_z} = \Big{(} C_{x,y,k_z;t\uparrow},  C_{x,y,k_z;t\downarrow},  C_{x,y,k_z;b\uparrow},  C_{x,y,k_z;b\downarrow} \Big{)}^T$, where $(x,y)$ labels the site on the effective square lattice and the lattice constant is $a$. 
Then, we consider the following superconducting Hamiltonian with on-site s-wave pairing,
\begin{align}
\begin{split}
	H_\text{SC} = \sum_{x,y} \Big{[} &\Delta_{x,y} \sum_{o\in\{t,b\}} \Big{(} C_{x,y,k_z;o\uparrow}^\dagger C_{x,y,-k_z;o\downarrow}^\dagger \\
	&-C_{x,y,k_z;o\downarrow}^\dagger C_{x,y,-k_z;o\uparrow}^\dagger \Big{)} +\text{H.c.} \Big{]}
\end{split}
\end{align}
where we set $\Delta_{x,y}= \Delta_0 \tanh (r/\xi_0) \exp(i\phi)$ in the limit of extreme type-II superconductor, here $r=\sqrt{x^2+y^2}$ and $\phi=\arctan\frac{y}{x}$. And we ignore both vector potential and Zeeman coupling arising from the magnetic field, since these two terms do not qualitatively change the results\cite{CKChiu-PRB-2011}. Therefore, we take the following BdG Hamiltonian for the vortex line,
\begin{align}\label{eq:total-BdG}
	H_\text{BdG}(k_z)= \begin{pmatrix}
	H_\text{TI}(k_z)-\mu & H_{\text{SC}} \\
	H_{\text{SC}}^\ast & -H_\text{TI}^*(-k_z)+\mu
	\end{pmatrix}
\end{align}
with the Nambu basis $\Big{(} \Psi_{x,y,k_z}, \Psi_{x,y,-k_z}^\dagger \Big{)}^{T}$. Then, we perform numerical calculation of the eigenvalues of BdG Hamiltonian in Eq.~\eqref{eq:total-BdG} on the effective square lattice with $a=1$ and $L_x=L_y=101$ (large enough). And $\Delta_0=0.1$ and $\xi_0=1$ are used.

We discuss the main results in the following subsections. In Subsection.~\ref{sub-sec-topo-range}, we focus on $k_z=0$ and $k_z=\pi$, and analyze the vortex line states as a function of chemical potential $\mu$. The bulk gap closing of vortex line indicates the occurrence of VPT. It helps to determine the relationship between topological range and strength of hexagonal warping term. In Subsection.~\ref{sub-sec-vlband}, we calculate the vortex line bands $E(k_z)$ at some particular chemical potential around $\mu_c$. In Subsection.~\ref{sub-sec-cons}, we discuss the consistence between the conclusion from Berry phase calculation and that from BdG calculation.

\subsection{Range for the Topological Phase}\label{sub-sec-topo-range}
In this subsection, we fix $k_z=0$ or $k_z=\pi$ to calculate the eigenvalues as a function of chemical potential. As for fully gapped topological systems, the band touchings happening in the time-reversal-invariant plane ($k_z=0$ and $k_z=\pi$) can alter the fermion number parity, resulting in the change of the $Z_2$ topological invariant defined for the vortex line. 

\begin{figure}[!htbp]
	\centering
	\includegraphics[width=3.4in]{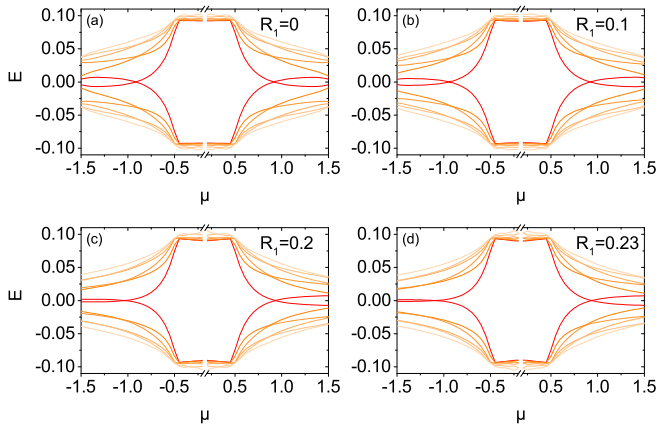}
	\caption{(color online) Several lower eigen-energies of the TI-SC system Eq.~\ref{eq:total-BdG} as a function of chemical potential $\mu$ in the $k_z=0$ plane. A $101\times 101$ square lattice is used in calculation. Different strength for hexagonal warping term are used: in (a) $R_1=0$; (b) $R_1=0.1$; (c) $R_1=0.2$; (d) $R_1=0.23$. In all cases $R_2=-2R_1$. The parameters for superconducting Hamiltonian are $\Delta_0=0.1$ and $\xi_0=1.0$, and other parameters for TI are the same as in Fig.~\ref{fig:BrPh-TI-TitBd}.}
	\label{fig:Eng-TISC-mu-hex}
\end{figure}

Firstly, we focus on $k_z=0$ case. From Fig.~\ref{fig:Eng-TISC-mu-hex}(a-d) (see the red solid lines), we can clearly see the gap closing phenomenon at two critical points: $\mu=\mu_c^{\pm}$, where $\mu_c^{+}$ ($\mu_c^{-}$) lays in the conduction (valence) band. And $\mu_c^{\pm}$ occur past the onset of bulk doping. More importantly, $\mu_c^{+}$ is almost independent of $R_1$; while $\mu_c^{-}$ is dramatically enlarged when $R_1$ is increased. In particular, our calculation shows that when $R_1=0$, $\mu_c^{-}=-0.91$, and when $R_1=0.23$, $\mu_c^{-}=-1.1$. 

\begin{figure}[!htbp]
	\centering
	\includegraphics[width=3.4in]{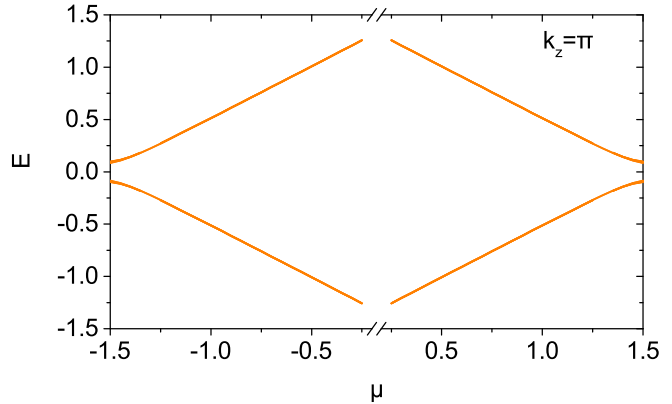}
	\caption{(color online) Several lower eigen-energies of the TI-SC system Eq.~\ref{eq:total-BdG} as a function of chemical potential $\mu$ in the $k_z=\pi$ plane. We only plot $R_1=R_2=0$ case here since the hexagonal warping doesn't change the energies obviously, and other parameter are the same as calculation of $k_z=0$ case in Fig.~\ref{fig:Eng-TISC-mu-hex}. There is a insulating gap in whole region. So VPT points are determined by $\mu_c^{\pm}$ in the $k_z=0$ plane.}
	\label{fig:Eng-TISC-mu-kzpi}
\end{figure}

Then, let us check if there exists gap closing in $k_z=\pi$ plane. The answer is ``No''. The results are shown in Fig.~\ref{fig:Eng-TISC-mu-kzpi} for $R_1=0$ case. Almost the same results are found when we increase $R_1$ (not shown). The reason is simple and explained as follows. According to our choice of parameters $M=B=-0.5$, thus in the $k_z=\pi$ plane, the single particle Hamiltonian is just a trivial insulator phase with a band gap 3.0 at $\Gamma$ point, since $\pm|M-B\cdot 2\left\lbrack 1-\cos(k_ia)\right\rbrack|=\pm 1.5$ at $(k_x,k_y)=(0,0)$ for any $R_1$. And we only focus on $\mu\in[-1.5,1.5]$ in our BdG calculation, which is just in the insulator gap. There is no in-gap states because of the absence of bulk Fermi surface. This explains no gap closing occurs in the $k_z=\pi$ plane. 

\begin{figure}[!htbp]
	\centering
	\includegraphics[width=3.4in]{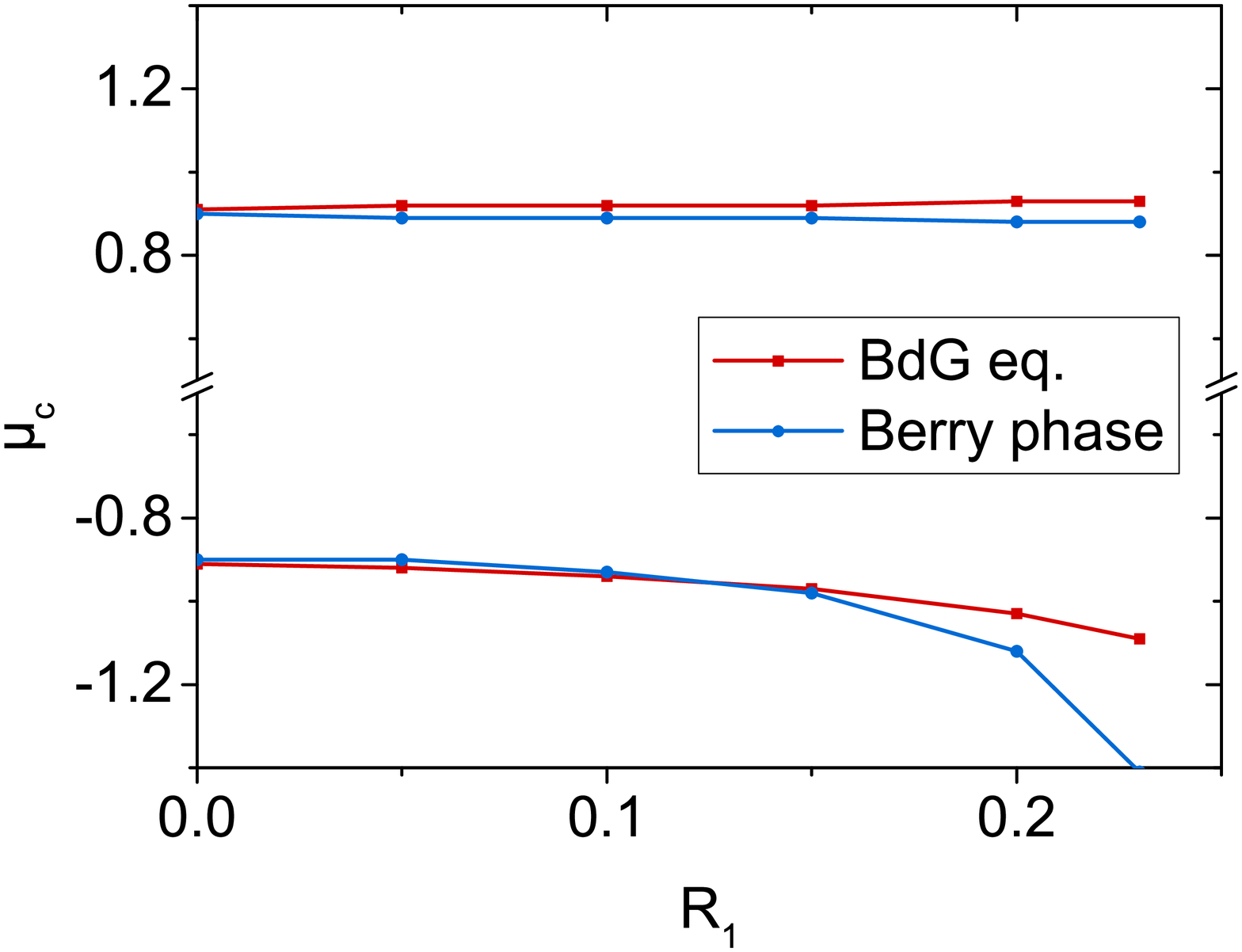}
	\caption{(color online) The VPT critical points calculated by full quantum mechanics of BdG equation in Fig.~\ref{fig:Eng-TISC-mu-hex} (blue) and by Fermi surface Berry phase in Fig.~\ref{fig:BrPh-TI-TitBd} (red) as variation of the hexagonal warping strength $R_1$. $R_2=-2R_1$ is kept and all other parameters are the same as original figures. These explicitly show that the hexagonal warping enlarges the topological region of the vortex line.}
	\label{fig:muc-BdG-BrPh}
\end{figure}

More explicitly, the VPT points $\mu_c^{\pm}$ as function of hexagonal warping strength, which are calculated by Berry phases and BdG equation from last and this section, is showed in Fig.~\ref{fig:muc-BdG-BrPh}. It is consistent that $\mu_c^-$ decreases as the $R_1$ increases while $\mu_c^+$ almost unaffected in both methods.
Therefore, $\mu_c^{\pm}$ determined in the $k_z=0$ plane determines the topological range $\delta\mu_c$, and it can be indeed enlarged as the increasing of $R_1$, thus stable MZMs at ends of vortex line could be well achieved in doped TIs.

\subsection{Vortex line bands}\label{sub-sec-vlband}
In this subsection, we show the low energy band dispersion of vortex line at different chemical potentials. 
As we discussed in Sec.~\ref{sec:BrPh}, to see weather it's 1D TSC phase which has fully gapped bulk except the VPT point,
it's necessary to calculate for vortex line bands as a function of $k_z$. Due to the particle-hole symmetry, only results for $k_z\ge0$ are calculated.

\begin{figure}[!htbp]
	\centering
	\includegraphics[width=3.4in]{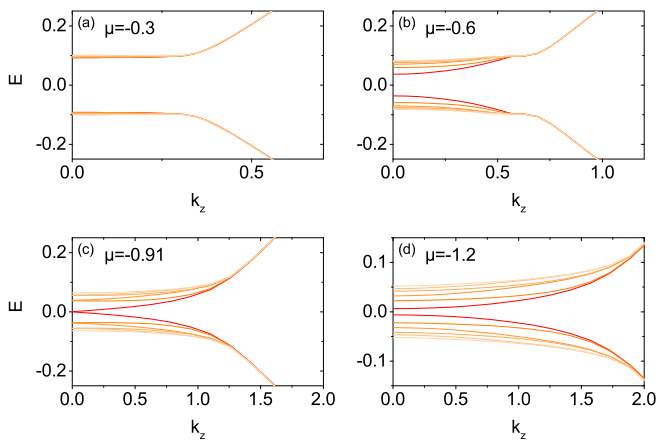}
	\caption{(color online) Several lower eigen-energies of the TI-SC system Eq.~\ref{eq:total-BdG} as a function of $k_z$ for $R_1=0$. Only results for $k_z\ge0$ are shown due to the particle-hole symmetry. And, different chemical potential $\mu$ are used: in (a) $\mu=-0.3>\mu_c^{-}\approx-0.91$; (b) $\mu=-0.6>\mu_c^{-}$; (c) $\mu=\mu_c^{-}$; and (d) $\mu=-1.2<\mu_c^{-}$. Other parameters are the same as in Fig.~\ref{fig:Eng-TISC-mu-hex}. We find that as $\mu$ decrease from -0.3 to -1.2, the energy gap closes at $\mu_c^{-}$ at $k_z=0$ and reopens after that.}
	\label{fig:Eng-TISC-kz-mu}
\end{figure}

\begin{figure}[!htbp]
	\centering
	\includegraphics[width=3.4in]{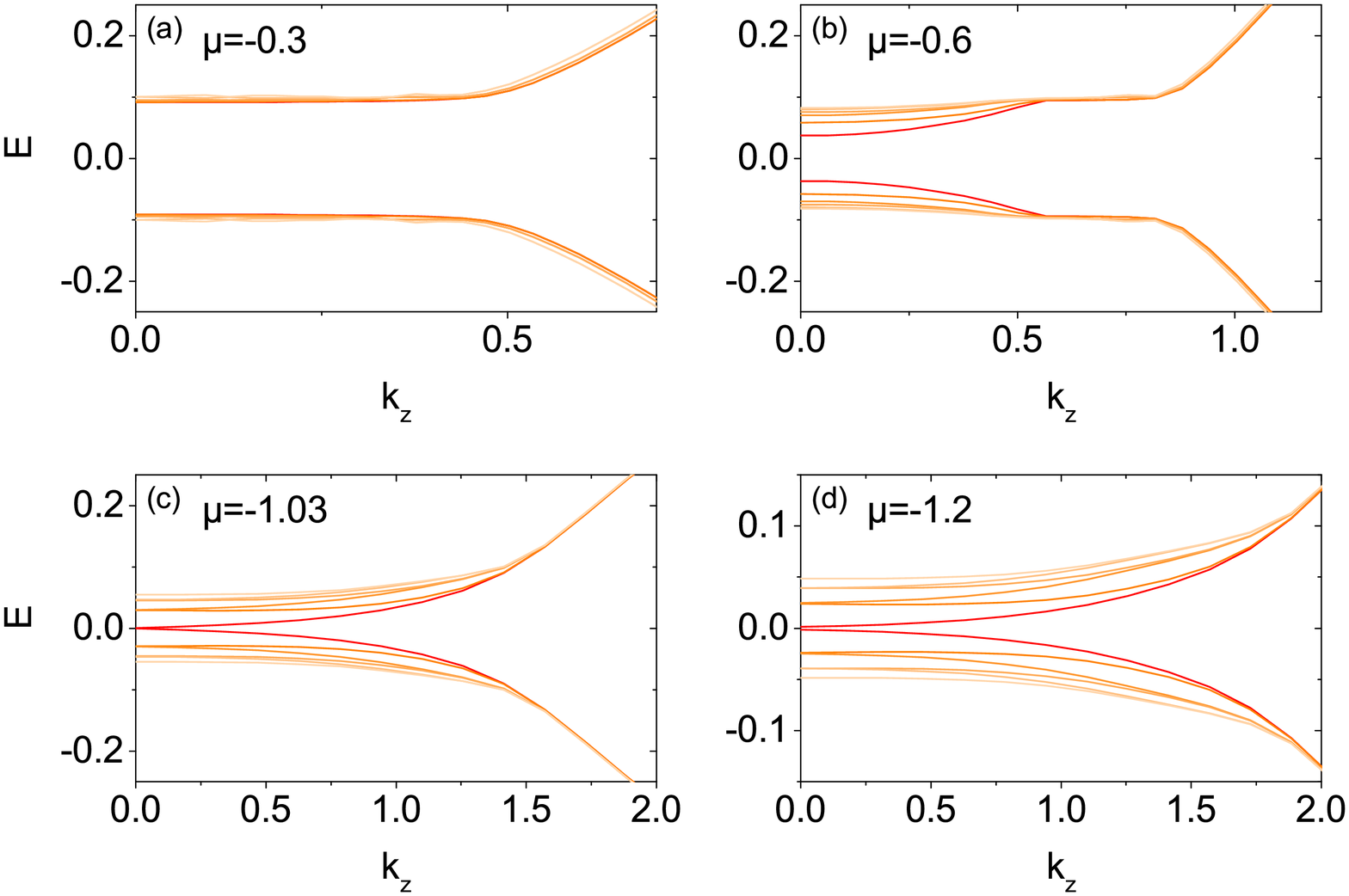}
	\caption{(color online) Several lower eigen-energies of the TI-SC system Eq.~\ref{eq:total-BdG} as a function of $k_z$ for $R_1=0.2$. Again, only results for $k_z\ge0$ are shown due to the particle-hole symmetry. And, different chemical potential $\mu$ are used: in (a) $\mu=-0.3>\mu_c^{-}\approx-1.03$; (b) $\mu=-0.6>\mu_c^{-}$; (c) $\mu=\mu_c^{-}$; and (d) $\mu=-1.2<\mu_c^{-}$. Other parameters are the same as in Fig.~\ref{fig:Eng-TISC-kz-mu}. Similar to the $R_1=0$ case, the energy gap closes at $\mu_c^{-}$ at $k_z=0$ and reopens when $\mu$ decreases from -0.3 to -1.2, so the hexagonal warping won't change the topology.}
	\label{fig:Eng-TISC-kz-mu--hex}
\end{figure}

For different $R_1$, similar results are found. Two cases are exhibited: (1) in Fig.~\ref{fig:Eng-TISC-kz-mu}, $R_1=0$ is used; and (2) in Fig.~\ref{fig:Eng-TISC-kz-mu--hex}, $R_1=0.23$ is used. Note that we only present the results for $k_z\ge0$ because of the particle-hole symmetry. Note that $\mu_c^{+}=-\mu_c^{-}$ for $R_1=0$ due to the presence of chiral symmetry. When $\mu>\mu_c^{-}$ in Fig.~\ref{fig:Eng-TISC-kz-mu}(a,b) and Fig.~\ref{fig:Eng-TISC-kz-mu--hex}(a,b), the vortex line is topological with nontrivial $Z_2$ topological invariant. In other words, the vortex line is a 1D topological superconductor. The gapless band dispersion can be found in Fig.~\ref{fig:Eng-TISC-kz-mu}(c) and Fig.~\ref{fig:Eng-TISC-kz-mu--hex}(c), and the band touching only occurs at $k_z=0$. It represents the topological phase transition. When we decrease the chemical potential to $\mu<\mu_c^{-}$, the vortex line is just trivial with a tiny gap proportional to $\Delta_0^2/\mu$. 
 
\subsection{Consistency with the semi-classical formula}\label{sub-sec-cons}	
In this subsection, we give a brief comparison between results in Sec.\ref{sec:BrPh} and Sec.\ref{sec:BdG-Calcu} and show they are qualitatively consistent.

From Fig.~\ref{fig:BrPh-TI-TitBd} in Sec.\ref{sec:BrPh}, we know the SU(2) Berry phase which enables us to estimate the energy of vortex bound states via the semi-classical formula in Eq.~\eqref{eq:vpt-condition}. At $\mu=\mu_c^{\pm}$ in the $k_z=0$ plane which gives rise to $\gamma(\mu_c^{\pm},0)=\pi$, we firstly find the vortex line becomes gapless and secondly notice that all the states have two-fold degeneracy. 

Then, let us look at the results in Fig.~\ref{fig:Eng-TISC-mu-hex} in Sec.\ref{sec:BdG-Calcu}, we also find that at $\mu=\mu_c^{\pm}$ in the $k_z=0$ plane, the lowest energy states are gapless and other higher energy states owns two-fold degeneracy. This is exactly consistent with the above results obtained from the semi-classical formula based on the calculation of SU(2) Berry phase.


\section{The relationship between VPT and topological surface states}\label{sec:Topo-Surf-Stt}
In this section, we discuss the relationship bettwen VPT and the disappearance of topological surface states (TSSs), by considering the TIs model with hexagonal warping terms. The disappearance of TSSs means the localized surface states emerges into the bulk and becomes extended at $\mu_{\text{TSS}}^\pm$, see the following Fig.~\ref{fig:dispersion-TSS}. 

\begin{figure}[!htbp]
	\centering
	\includegraphics[width=3.3in]{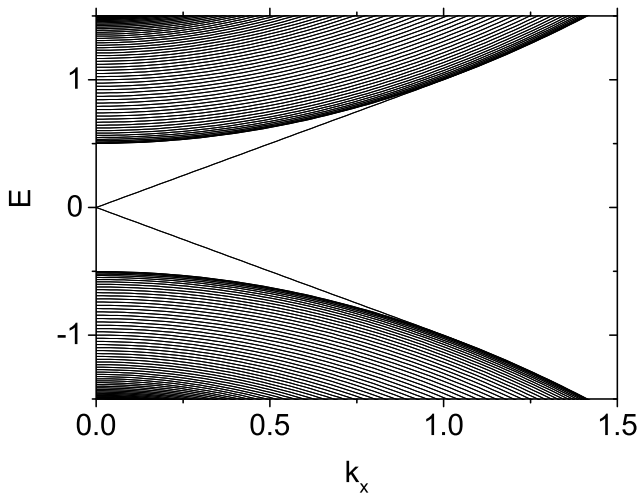}
	\caption{Dispersions of TI's bulk states and topological surface states (TSSs) at $k_y=0$. This can be plotted by transform $z$-direction in Eq.~\eqref{eq:TI-Conti} into real space with open boundary conditions. We use $N_z=60$ sites in $z$-direction with effective lattice constant $a=1$, and other parameters are the same as Fig.~\ref{fig:Disp2-TI-Conti} except the absence of hexagonal warping. We find that the TSSs coexists with bulk state until the chemical potential approaches $\mu_{\text{TSS}}^{\pm}$, and the energies are distinguishable before that.}
	\label{fig:dispersion-TSS}
\end{figure}

As we know, there is no bulk Fermi surface when $\vert\mu\vert<\vert M\vert$, then what is the reason why the vortex line is topological? It is because of the existence of topological surface states. Inside the vortex on top and bottom surface, there exists MZMs\cite{LFu-PRL-2008}. Strong evidence has shed on the observation of MZMs\cite{hTC-PRB-2016,JFJia-PRL-2016}. Moreover, when $\vert M\vert<\mu<\mu_{\text{TSS}}^{+}$ or $\mu_{\text{TSS}}^{-}<\mu<-\vert M\vert$, the topological surface still survives with coexistence of bulk states. Furthermore, when $\mu>\mu_{\text{TSS}}^{+}$ or $\mu<\mu_{\text{TSS}}^{-}$, the localized surface state eventually disappears and evolutes into the bulk that becomes extended states along the z direction.
Then, we can ask the following important question: what is the relationship between $\mu_{\text{TSS}}^{\pm}$ and $\mu_c^{\pm}$?
 
Firstly, let us describe how to determine $\mu_{\text{TSS}}^{\pm}$ for the disappearance of TSSs. 
In the absence of hexagonal warping term, i.e., $R_1=0$, we learn that the SU(2) Berry phase equals to $\pi$ when $M-Bk^2=0$ that gives rise to $\mu_c^{+}=v\sqrt{M/B}$ in the $k_z=0$ plane\cite{Vishwanath-PRL-2011}. Because it resembles two copies of a TI surface of which the Berry phase is exactly $\pi$. In the other hand, if we focus on $k_x=0$ and take open boundary condition along z direction. Note $k_y$ is a good quantum number. The Hamiltonian $H(z,-i\partial_z)$ reads,
\begin{align}
\begin{split}
   H(z,-i\partial_z) & = H_{1D} + vk_y\alpha_y \\
   H_{1D} &= (M-Bk_y^2+B\partial_z^2)\beta_y -iv\partial_z\alpha_z 
\end{split}
\end{align}
In the spirit of $\mathbf{k}\cdot \mathbf{p}$ approximation, we firstly treat $vk_y\alpha_y$ as a perturbation term. As for $H_{1D}$, we know it have localized state with zero energy once $M-Bk_y^2\ge0$ is satisfied, so that $H_{1D} \vert\psi(z)\rangle = 0$. The dispersion of the topological surface states is given by $E_{TSS}(k_y)=vk_y\int dz \langle \psi(z)\vert \alpha_y\vert\psi(z)\rangle$. Therefore, we learn that $\mu_{\text{TSS}}^{\pm} = \pm v\sqrt{M/B}$. Surprisingly, we find $\mu_{\text{TSS}}^{\pm}=\mu_c^{\pm}$ in the absence of hexagonal warping term ($R_1=0$). 

\begin{figure}[!htbp]
	\centering
	\includegraphics[width=3.4in]{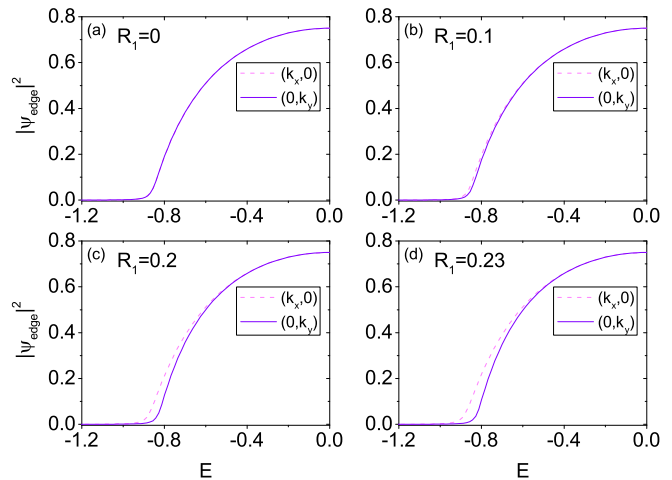}
	\caption{(color online) Calculated TSSs' wave function at edge $\vert\Psi_{\text{edge}}\vert^2$ defined by Eq.~\eqref{eq:edge-state-wf} as a function of energy $E$ in TI TB model Eq.~\eqref{eq:TI-TitBd} with open boundary condition in $z$-direction. The dashed light magenta line is for $(k_x,0)$, while the solid violet line is for $(0,k_y)$. And different $R_1$ are used: in (a) $R_1=0$; (b) $R_1=0.1$; (c) $R_1=0.2$ and (d) $R_1=0.23$. Other parameters are the same as Fig.~\ref{fig:dispersion-TSS}. We find that the TSSs roughly disappear at $E\sim-0.9$ and is almost unaffected by hexagonal warping, which is not coincident with the VPT $\mu_c^-$.}
	\label{fig:WFEdge-Eng-TitBd}
\end{figure}

We next carry out a numerical calculation to search for the disappearance of TSSs. Based on the tight-binding model of TIs in Eq.~\eqref{eq:TI-TitBd}, we perform numerical calculation on a slab configuration along z direction ($z\in[1,L_z]$). To determine $\mu_{\text{TSS}}^{\pm}$, we define 
\begin{align}\label{eq:edge-state-wf}
\begin{split}
	\vert\Psi_{\text{edge}}\vert^2 =
	\max\{ &\left\vert \langle \Psi(z=L_z) \vert \Psi(z=L_z) \rangle \right\vert, \\
	&\left\vert \langle \Psi(z=1) \vert \Psi(z=1) \rangle \right\vert \} 
\end{split}
\end{align}
to measure the disappearance of TSSs. The results are shown in Fig.~\ref{fig:WFEdge-Eng-TitBd}. In the absence of hexagonal warping term in Fig.~\ref{fig:WFEdge-Eng-TitBd}(a), we firstly confirm $\mu_{\text{TSS}}^{\pm}$ equals to $\mu_c^{\pm}$, consistent with the above analytical analysis.

Next, we consider the hexagonal warping effect. In Fig.~\ref{fig:WFEdge-Eng-TitBd}(b-d), we increase $R_1$ from $0.1$ to $0.23$, we find $\mu_{\text{TSS}}^{-}$ is decreased. For example, when $R_1=0.23$ we find $\mu_{\text{TSS}}^{-}=-0.85$. Recall that $\mu_c^{-}= -1.1$ from the BdG calculation (see Fig.~\ref{fig:Eng-TISC-mu-hex}(d)), therefore, we find $\mu_c^{-}<\mu_{\text{TSS}}^{-}<0$. It means the vortex line is still topological when $\mu_c^{-}<\mu<\mu_{\text{TSS}}^{-}$ even when the TSSs have been already disappeared. Thus, in that case, the MZMs at ends of vortex line merely stems from the bulk Fermi surface of TIs.

In brief, the occurrence of the VPT is independent on the disappearance of TSSs, namely, $\mu_{\text{TSS}}^{\pm} $ are usually not equal to $\mu_{c}^{\pm}$. Our theoretical results may help to explain experiment.

\section{Summary}\label{sec:Conclusion}
In this paper, we have applied semiclassical formula to analytically calculate the vortex phase transition in a superconducting vortex line by including a hexagonal warping term. The hexagonal warping term enlarges the region of the chemical potential for the topological vortex line. It is interesting to note that the topological vortex line and its associated Majorana zero modes may exist as the chemical potential is deep inside the bulk band of the topological insulator so that topological surface state is absent.

%

\section*{Acknowledgments}
We would like to thank Shengshan Qin, Congcong Le and P. Hosur for numerous enlightening discussions. LHH is supported by AFOSR FA9550-14-1-0168. FCZ is supported by NSFC grant 11674278, National Basic Research Program of China (No. 2014CB921203), and the CAS Center for Excellence in Topological Quantum Computation.

\appendix
\section{Calculations for the Continuous Model}\label{sec:appendix-berry-phase}
\begin{figure}[!htbp]
	\centering
	\includegraphics[width=3.4in]{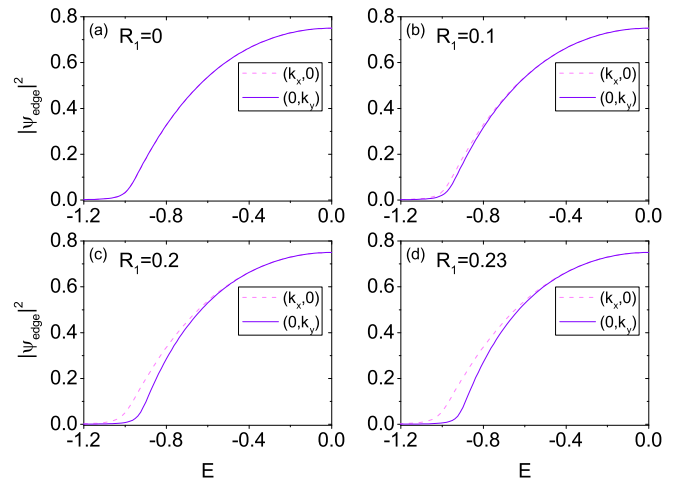}
	\caption{(color online) Calculated TSSs' wave function at edge $\vert\Psi_{\text{edge}}\vert^2$ defined by Eq.~\eqref{eq:edge-state-wf} as a function of energy $E$ in TI continuous model Eq.~\eqref{eq:TI-TitBd} with open boundary condition in $z$-direction. The parameters are the same as Eq.~\eqref{eq:total-ham}. Similarly, the TSSs roughly disappears at $E\sim-0.9$.}
	\label{Appendix-fig:WFEdge-Eng-Conti}
\end{figure}

In this appendix, we carry out the calculation based on the continuous model to show the consistency with the results from lattice model (shown in the main text).
We calculate $\vert\Psi_{\text{edge}}\vert^2$ from Eq.~\eqref{eq:edge-state-wf} as a function of energy $E$. The results are shown in Fig.~\ref{Appendix-fig:WFEdge-Eng-Conti}, which is consistent with the results from lattice model (see Fig.~\ref{fig:WFEdge-Eng-TitBd} in the main text).

\bibliographystyle{apsrev4-1}
\bibliography{ref}	
	
\end{document}